\documentclass[letterpaper,aps,twocolumn,showpacs,superscriptaddress,floats,prl]{revtex4}

\usepackage{amsmath}
\usepackage{graphicx,epsfig,dsfont}
\usepackage{amssymb}

\newcommand{\nn}{\nonumber}

\begin{document}

\title{Mott metal-insulator transition on compressible lattices}

\author{Mario Zacharias}
\affiliation{Institut f\"ur Theoretische Physik, Universit\"at zu K\"oln,
Z\"ulpicher Str. 77, 50937 K\"oln, Germany
}
\author{Lorenz Bartosch}
\affiliation{Institut f\"ur Theoretische Physik, Goethe-Universit\"at, 60438 Frankfurt am Main, Germany}
\author{Markus Garst}
\affiliation{Institut f\"ur Theoretische Physik, Universit\"at zu K\"oln,
Z\"ulpicher Str. 77, 50937 K\"oln, Germany
}

\begin{abstract}
The critical properties of the finite temperature Mott endpoint are drastically altered by
a coupling to crystal elasticity, i.e.,
whenever it is amenable to pressure tuning. Similar as for critical piezoelectric 
ferroelectrics, the Ising criticality of the electronic system is preempted by an isostructural instability, and long-range shear forces suppress microscopic fluctuations. As a result, the endpoint is governed by Landau criticality. Its hallmark is thus a breakdown of Hooke's law of elasticity 
with a non-linear strain-stress relation 
characterized by a mean-field exponent. 
Based on a quantitative estimate, we predict critical elasticity to dominate the temperature range $\Delta T^*/T_c \simeq 8\%$ close to  the Mott endpoint of  $\kappa$-(BEDT-TTF)$_2$X. 
\end{abstract}

\date{\today}

\maketitle

\begin{figure}
\centering
\includegraphics[scale=0.4]{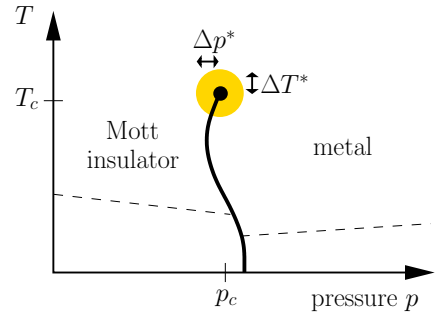}
\caption{
The line of first-order Mott transitions terminates at a finite temperature critical endpoint. 
The pressure-tuned endpoint exhibits Landau criticality due to the non-perturbative Mott-elastic coupling (yellow regime).
The dashed lines bound possible low temperature phases like antiferromagnetism and superconductivity.
}
\label{fig:PDkappa}
\end{figure}

Strong repulsion between electrons in a solid containing approximately one electron per lattice site promotes insulating behavior as the electrons' motion is inhibited by the large energetic cost of having a site doubly occupied. A so-called Mott insulator is favored if the on-site Coulomb repulsion $U$ exceeds the kinetic energy $W$ while metallic behavior prevails for $U/W \ll 1$, allowing for a first-order metal-insulator transition at a critical ratio of $U/W$. Usually, this critical ratio can be controlled by an external control parameter like pressure, $p$, or doping, giving rise to a line of first-order transitions in the $(p,T)$ phase diagram plane, where $T$ is temperature. This line of transitions terminates in a second-order critical endpoint at a finite 
temperature $T_c$ beyond which the system can be smoothly transformed from the insulating to the metallic regime by varying $p$ and $T$, see Fig.~\ref{fig:PDkappa}. 

The nature of the finite-$T$ critical endpoint of the Mott transition attracted some attention recently. From general considerations, one expects it to belong to the Ising universality class \cite{Castellani79,Kotliar00} similarly to the endpoint of the liquid-gas transition. The double occupancy of a single site plays here the role of the local Ising order parameter of the transition. Measurements of the electrical conductivity on Cr-doped V$_2$O$_3$ \cite{Limelette03} have confirmed this expectation. However, transport \cite{Kagawa05} and NMR \cite{Kagawa09} measurements on the quasi two-dimensional organic charge-transfer salt $\kappa$-(BEDT-TTF)$_2$X \cite{Lefebvre00,Limelette03b,Fournier03,ToyotaBook} questioned this interpretation and suggested a different universality class. 
Subsequently, various theories were proposed to account for the unconventional behavior \cite{Imada05,Imada10,Sentef11,Semon12}.
In particular, it was demonstrated \cite{Papanikolaou08} that the   
analysis of the conductivity is intricate as its scaling exponents are not necessarily directly related to the scaling dimension of the Ising order parameter. Taking this into account,  the conductivity experiments could be reconciled with Ising universality. 

In contrast to transport quantities, thermodynamics   
in principle allows for a straightforward interpretation in terms of a standard critical scaling analysis. Ultrasound experiments on $\kappa$-(BEDT-TTF)$_2$X revealed a pronounced softening close to the Mott endpoint \cite{Fournier03}, 
but an analysis of the critical behavior has not been performed yet.  Thermal expansion measurements \cite{deSouza07,Bartosch10} on the same system are consistent with two-dimensional critical Ising behavior even though an experimental verification of scaling exponents was not possible until now.

While the Mott transition of an idealized correlated electron system should exhibit Ising criticality, a coupling of electrons to crystal elasticity drastically changes its critical properties. 
The Mott transition is very sensitive to the presence of elastic strain in the atomic crystal lattice as it alters the overlap integrals of electron wavefunctions between adjacent lattice sites. This in turn changes the kinetic energy $W$, allowing for an efficient tuning of the transition by applying external stress, e.g., a compressive pressure as in Fig.~\ref{fig:PDkappa}.
Conversely, the critical Mott system exerts an internal pressure on the elastic system to which the crystal lattice responds. 
The detection of this response with the help of dilatometric measurements, e.g., thermal expansion, is a convenient and sensitive probe of Mott criticality. Sufficiently far away from criticality, this lattice response is perturbative and the critical behavior itself remains unaffected. 

However, close to the Mott endpoint, the lattice necessarily reacts in a non-perturbative manner to the internal stress, leading to a vanishing elastic modulus and thus to a breakdown of Hooke's law of elasticity. This was noted before by Krishnamurthy and collaborators \cite{Majumdar94,Hassan05} within the framework of the compressible Hubbard model. Importantly, we point out here that 
this breakdown of Hooke's law is generically accompanied by a crossover 
from Ising 
criticality to Landau critical behavior
with mean-field exponents. 
At the origin of this change of universality class are the long-ranged shear forces of the atomic crystal lattice. They become instrumental as an elastic modulus becomes small so that eventually Landau mean-field behavior prevails close to the Mott endpoint.

Within an effective field theoretic description, we consider a coupling of the elastic strain tensor $\varepsilon_{ij}$ to the Ising order parameter $\phi$ of the Mott transition,
\begin{align}
\mathcal{L}_{\rm int} = - \gamma_{1,ij} \varepsilon_{ij} \phi + \frac{1}{2} \gamma_{2,ij} \varepsilon_{ij} \phi^2,
\end{align}
where $\gamma_{1,ij}$ and $\gamma_{2,ij}$ are elastic coupling tensors. 
Interestingly, as the Ising symmetry of the Mott endpoint is an emergent symmetry, a linear coupling $\gamma_{1,ij}$ of strain to the order parameter is generally allowed. The quadratic coupling $\gamma_{2,ij}$ is less important, but for completeness we include it in the following discussion.
The effect of a linear coupling of an order parameter to strain was considered by Levanyuk  and Sobyanin \cite{Levanyuk70} and independently by Villain \cite{Villain70} in the context of critical ferroelectrics, who showed that it suppresses critical long-wavelength fluctuations. As a consequence, for sufficiently large $\gamma_{1,ij}$, the Ginzburg criterion is never fullfilled, thus stabilizing Landau mean-field behavior.

An analysis of the effective elastic Hamiltonian 
\cite{Cowley76,Folk76} suggests that even if the critical subsystem is controlled by an interacting renormalization group fixed point, a small linear elastic coupling $\gamma_{1,ij}$ can recover mean-field behavior sufficiently close to the transition.
It turns out that the singularities associated with the Mott endpoint induce via the coupling $\gamma_{1,ij}$ a macroscopic instability of the crystal lattice. At such a lattice instability, an elastic modulus associated with the macroscopic strain field $E_{ij}$, i.e., an eigenvalue of the $6\times 6$ elastic constant matrix, $C_{\rho \nu} \cong C_{ijkl}$, vanishes \cite{BornBook}. 
In addition to $E_{ij}$, the elastic strain also contains a part that carries finite momentum and describes the long-wavelength acoustic modes 
\begin{align}
\varepsilon_{ij}({\bf r}) = E_{ij} +  e_{ij}({\bf r}),
\end{align}
with $\int d^3 r\, e_{ij}({\bf r}) = 0$. Importantly, at a lattice instability, the velocities of the acoustic modes soften, but generally remain finite due to the shear stiffness of the solid. 
The phonon velocities are determined by the $3\times 3$ matrix $M_{ik}({\bf q}) = \sum_{jl} C_{ijkl} q_j q_l$ depending on momentum $\bf q$, and its eigenvalues generally remain positive even if an eigenvalue of $C_{ijkl}$ vanishes. 
Possible exceptions may be acoustic modes with momenta in certain lattice directions. For general momenta, however, the acoustic modes remain non-critical and, as a consequence, the structural transition is described by Landau's mean-field theory \cite{Cowley76,Folk76}. 

Neglecting these non-critical acoustic modes, the macroscopic strain $E_{ij}$ is determined by the effective potential
\begin{align}
\mathcal{V}(E_{ij}) &= \frac{1}{2} E_{ij} C^{(0)}_{ijkl} E_{kl} + E_{ij} \sigma_{ij} 
\nn\\
&
\qquad+ f_{\rm sing}(t_0 + \gamma_{2,ij} E_{ij}, h_0 + \gamma_{1,ij} E_{ij}),
\end{align} 
where $C^{(0)}_{ijkl}$ is the elastic constant matrix in the absence of Mott-elastic couplings $\gamma_{n,ij}$, with $n=1,2$, and $\sigma_{ij}$ is an externally applied macroscopic stress. The free energy density $f_{\rm sing}$ is attributed to the 
critical electronic subsystem and is governed by the Ising universality class. The two relevant perturbations $h_0$ and $t_0$ quantify the distance to criticality (for $\gamma_{n,ij} = 0$) and generally depend on temperature $T$. 
In order not to distract with cumbersome notation and to focus on the mechanism at play, let us assume that the electronic subsystem mainly couples to a certain singlet, $E$, of the irreducible representations of the crystal group. We can then limit ourselves to an effective potential for $E$ only,
\begin{align} \label{PotE}
\mathcal{V}(E) = \frac{K_0}{2} E^2  - E p + f_{\rm sing}(t_0+\gamma_2 E, h_0 + \gamma_1 E).
\end{align} 
Here $K_0$ is the corresponding modulus for $\gamma_n=0$. Furthermore, we considered for simplicity the application of a hydrostatic pressure, $\sigma_{ij} = - p \delta_{ij}$, assuming a finite overlap with the singlet $E$. The thermodynamic free energy density obtains after minimizing this potential with respect to $E$.

The sensitivity of the Mott endpoint with respect to pressure tuning becomes manifest if the potential \eqref{PotE} is minimized perturbatively in $\gamma_n$. In zeroth order one has $E = p/K_0$ so that the free energy density becomes
\begin{align}  \label{PertF}
\mathcal{F}_{\rm pert} = -\frac{p^2}{2 K_0} + f_{\rm sing}(t_0+\gamma_2 p/K_0, h_0 + \gamma_1 p/K_0).
\end{align} 
The elastic coupling induces a pressure dependence of the arguments of the function $f_{\rm sing}$  which allows to control the distance to criticality by varying $p$, thus enabling pressure-tuning of the Mott transition.

\begin{figure}
\centering
\includegraphics[width=0.9\columnwidth]{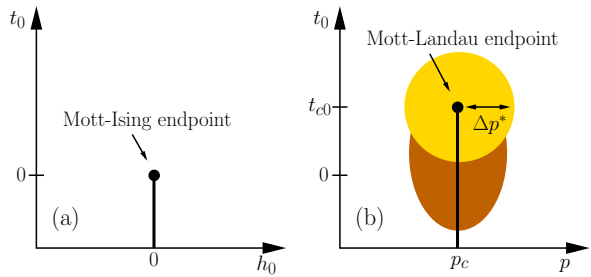}
\caption{Phase diagram close to the finite-$T$ Mott endpoint 
in the $(a)$ absence and $(b)$ presence of a linear elastic coupling $\gamma_1 \neq 0$ (and $\gamma_2=0$); $t_0$ and $h_0$ are the two relevant fields and $p$ is the pressure. Whereas for $(a)$ the endpoint is Ising critical for $(b)$ a crossover is induced from Ising to Landau mean-field criticality for $|t_0-t_{c0}| \lesssim t_{c0}$ and $|p-p_c| \lesssim \Delta p^*$.
In the uncolored and yellow shaded regime in $(b)$ thermodynamics is governed by Eq.~\eqref{PertF} and Eq.~\eqref{PotE2}, respectively, 
and in the brown shaded regime the full potential Eq.~\eqref{PotE} must be used.
}
\label{fig:PD}
\end{figure}

However, it is important to realize that such a perturbative treatment necessarily breaks down sufficiently close to the endpoint. This becomes evident after expanding the potential in a Taylor series, 
\begin{align} \label{PotE2}
\lefteqn{\mathcal{V}(E) = f_{\rm sing}(\bar t, \bar h) }\nn\\
&  -  (p-\bar p) \delta E + \frac{K}{2} \delta E^2  + \frac{u}{4!} \delta E^4  + \mathcal{O}(\delta E^5),
\end{align} 
where $\delta E = E - \bar E$, $\bar t = t_0+\gamma_2 \bar E$ and $\bar h =  h_0 + \gamma_1 \bar E$. The value of $\bar E$ is conveniently chosen such that the prefactor of the cubic term, $\delta E^3$, in the expansion just vanishes. The pressure $\bar p$ reads $\bar p = K_0 \bar E + (\gamma_2 \partial_{\bar t}+\gamma_1 \partial_{\bar h} ) f_{\rm sing}(\bar t, \bar h)$, and the quartic coupling $u$ is given by fourth-order derivatives of $f_{\rm sing}$. Importantly, the modulus gets renormalized by the susceptibilities $\chi_{a b} = -  \partial_{a}\partial_b f_{\rm sing}(\bar t, \bar h)$ with $a,b = \bar t, \bar h$,
\begin{align} \label{RenModulus}
K = K_0 - \gamma_1^2  \chi_{\bar h \bar h} - 2 \gamma_1 \gamma_2 \chi_{\bar  h \bar t}- \gamma_2^2  \chi_{\bar t \bar t}.
\end{align}
The most singular susceptibility is $\chi_{\bar h \bar h}$ which necessarily diverges, $\chi_{\bar h \bar h} \to \infty$, as the endpoint $(\bar t, \bar h) = 0$ is approached. Hence, irrespective of the magnitude of the linear elastic coupling, $\gamma_1\neq 0$, the divergence of $\chi_{\bar h \bar h}$ will drive the effective modulus to zero at a {\it finite} value of $\bar t$ where the Taylor expansion in Eq.~\eqref{PotE2} is well defined. The resulting isostructural instability at $K=0$ and $p=\bar p$ identifies a mean-field endpoint in the phase-diagram. This critical endpoint at $(t_{c0},p_c)$ preempts the Mott-Ising endpoint in Fig.~\ref{fig:PD}(b).
The coupled Mott-elastic system thus avoids the Ising singularities by developing a non-perturbative strain response to pressure changes. Minimization of Eq.~\eqref{PotE2} for $K=0$ yields $\delta E = (6 (p-\bar p)/u)^{1/\delta}$ with the Landau value $\delta = 3$,
clearly violating Hooke's law of elasticity. This violation sets in for $|p-\bar p| \lesssim \Delta p^*$  at $K=0$ with $\Delta p^* = K_0^{3/2}\sqrt{6/u}$.
 
\begin{figure}
\centering
\includegraphics[width=0.8\columnwidth]{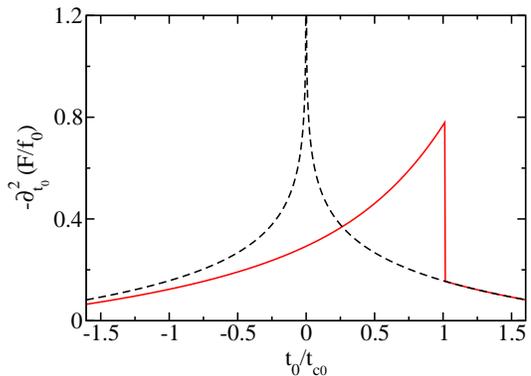}
\caption{The second derivative $- \partial^2_{t_0} \mathcal{F}$ as a function of $t_0$ exhibits a pure mean-field jump at criticality due to the linear Mott-elastic coupling (solid curve) that preempts the Ising singularity (dashed curve), see text.}
\label{fig:MFJump}
\end{figure} 

As a concrete example, we assume that the critical electronic subsystem is effectively two-dimensional, and the function $f_{\rm sing}$ in Eq.~\eqref{PotE} is determined by the 2d Ising model \cite{Fonseca03}
\begin{align} \label{2dIsing}
f_{\rm sing}(t,h)= f_0 \left(\frac{t^2}{8\pi} \log t^2 + |h|^{16/15} \Phi(t |h|^{-8/15}) \right),
\end{align}
where $f_0$ has the dimension of a free energy density (and $t$ and $h$ are assumed to be dimensionless). 
Using the results of Ref.~\cite{Fonseca03}, the scaling function $\Phi$ can be evaluated numerically. In order to illustrate the mean-field character of the shifted Mott endpoint, we show in Fig.~\ref{fig:MFJump} the second derivative $- \partial^2_{t_0} \mathcal{F}$ of the free energy density $\mathcal F$ for $\gamma_1 \neq 0$ (and $\gamma_2 = 0$). The pressure is fixed to the critical value $p_c$ so that the endpoint is crossed as a function of $t_0$, i.e., along the vertical axis in Fig.~\ref{fig:PD}(b). The solid curve shows the behavior obtained from minimizing the full potential \eqref{PotE} while the dashed curve follows from the perturbative expression \eqref{PertF}. The latter exhibits the characteristic logarithmic divergence of the 2d Ising model at $t_{0} = 0$. However, the non-perturbative renormalization of the elastic constant results in a preemptive mean-field transition at $t_{0c} > 0$ so that the logarithmic divergence is cut off and $- \partial^2_{t_0} \mathcal{F}$ instead shows a mean-field jump and remains finite \cite{Levanyuk70}. 

These considerations are directly relevant for $\kappa$-(BEDT-TTF)$_2$X close to its Mott endpoint.
In Ref.~\cite{Bartosch10}, the perturbative free energy density \eqref{PertF} was used together with Eq.~\eqref{2dIsing} for the interpretation of thermal expansion measurements. For the so-called $d8$-Br crystal $\#1$ in Ref.~\cite{Bartosch10}, for which $p-p_c \approx 50$\,bar at ambient pressure and $T_c \approx 30$\,K, the following fitting parameters were obtained: 
$f_0 \approx 5.7$\,bar, $h_0 + \gamma_1 p_c/K_0 \approx - 0.004 (T-T_c)/T_c$, 
and $\gamma_1/K_0 \approx 0.07/$\,kbar, where the scaling freedom was exploited to choose $t_0 + \gamma_2 p_c/K_0 = (T-T_c)/T_c$.
As the exact critical temperature $T_c$ of the crystal is not known, there is no reliable estimate for $\gamma_2$. In the following, we neglect the subleading corrections due to $\gamma_2$ and use $\gamma_2 = 0$.
A crucial question concerns the extension of the non-perturbative regime
in order to assess whether an experimental investigation of the crossover from Ising to Landau criticality is feasible. With the above fitting parameters and the estimate for the bare modulus $K_0 \approx 122$\,kbar \cite{Hassan05} we can estimate the width of the Landau critical regime in pressure, $\Delta p^*$, and temperature, $\Delta T^* \equiv t_{c0} T_c$ for this compound, see Fig.~\ref{fig:PDkappa},  
\begin{align} \label{XoverScales}
\Delta p^* \approx 45\,{\rm bar}
\qquad
\Delta T^* \approx 2.5\,{\rm K}.
\end{align}  
These values are sufficiently large to allow for an experimental detection of the crossover phenomena. In fact, the $d8$-Br crystal seems to be located already within the crossover regime as $\Delta p^*$ is on the same order as the distance $p-p_c$ \cite{comment}.

 \begin{figure}
\centering
\includegraphics[width=0.8\columnwidth]{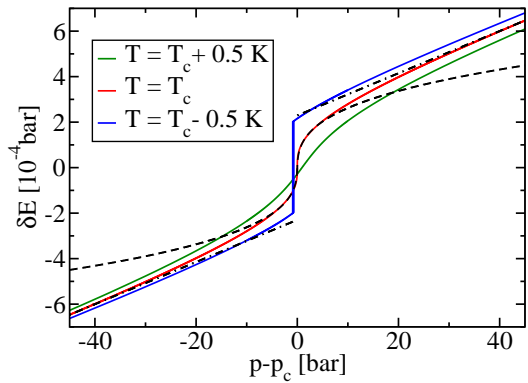}
\caption{Lattice strain $\delta E$ as a function of applied pressure close to the critical temperature $T_c$.
At $T_c$ (red line), the strain is linear in the applied pressure for $|p-p_c| \gtrsim \Delta p^*$ (dashed-dotted line asymptote)
but becomes non-linear close to the endpoint, where $\delta E \sim |p-p_c|^{1/\delta}$ with the mean-field exponent $\delta = 3$ (dashed line asymptote), signaling a breakdown of Hooke's law.
}
\label{fig:strain}
\end{figure} 

With the above fitting values and the value for $K_0$ we can predict the thermodynamics with the help of the potential \eqref{PotE} and Eq.~\eqref{2dIsing}.
In particular, the crossover at $\Delta p^*$ is illustrated in Fig.~\ref{fig:strain} which shows the expected lattice strain as a function of applied pressure.
Far away from the transition, $|p-p_c| \gg \Delta p^*$, the strain is linear in the applied pressure, thus obeying Hooke's law. However, at $T_c$ the pressure-strain relation becomes non-linear for $p\to p_c$ with mean-field exponent $\delta = 3$. This breakdown of Hooke's law and the concomitant divergence of the associated modulus serves as a smoking-gun criterion for the detection of the Landau critical regime where the Mott-elastic coupling becomes non-perturbative.

\begin{figure}
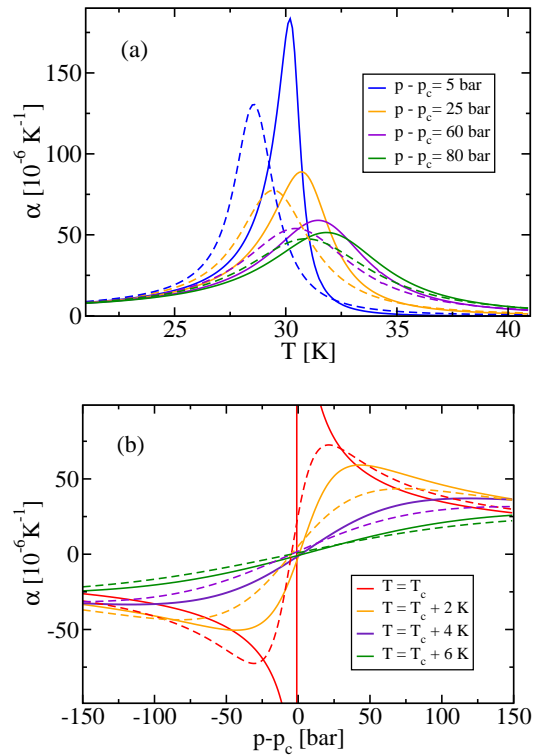

\centering
\includegraphics[width=0.8\columnwidth]{figure5}
\\[1.5em]
\includegraphics[width=0.8\columnwidth]{figure6}
\caption{Singular part of thermal expansion (solid line) as a function of (a) temperature for different pressure values and of (b) pressure for different temperatures using the estimate $T_c \approx 30$\,K. The perturbative behavior (dashed line) is a good approximation away from the endpoint, $|p-p_c| \gg \Delta p^*$ and $|T-T_c| \gg \Delta T^*$. 
}
\label{fig:Alpha}
\end{figure} 

In Fig.~\ref{fig:Alpha}(a) we show the thermal expansion, $\alpha = \partial_p\partial_T \mathcal{F}$, as a function of temperature for different pressure values (solid lines). For comparison, the dashed lines demonstrate the corresponding Ising critical behavior obtained from the perturbative expression of Eq.~\eqref{PertF}. The latter is a good approximation far away from the endpoint but fails close to it and, in particular, exhibits a peak at a temperature $\sim T_c - \Delta T^*$, that is smaller than $T_c$, see also Fig.~\ref{fig:PD}. The crossover is identified when the solid and dashed curves at a given pressure start to deviate substantially.  Finally, Fig.~\ref{fig:Alpha}(b) displays the thermal expansion as a function of pressure for different temperatures. 
Note that the sign change of the thermal expansion in Fig.~\ref{fig:Alpha}(b) can be related to entropy accumulation similarly as in the case of quantum criticality \cite{Garst05}. 
 
In Ref.~\cite{Papanikolaou08} the conductivity, $\sigma$,  was interpreted to scale with the energy-density of the Ising model, $\sigma \sim \partial_{t_0} f_{\rm cr}$. If this interpretation holds across the crossover discussed here, i.e., $\sigma \sim \partial_{t_0} \mathcal{F}$, one would also expect signatures in transport at scales $\Delta T^*$ and $\Delta p^*$ of Eq.~\eqref{XoverScales}. 
Interestingly, whereas the pressure dependence of $\sigma$ measured in Ref.~\cite{Kagawa05} does not show such a signature, there are indications of a crossover in $\sigma(T)$ at around $T_c \pm 1$\,K.
Clearly, detailed dilatometric studies are favored to identify unambiguously the crossover 
to Mott-Landau criticality. 
Our estimate for $\kappa$-(BEDT-TTF)$_2$X, Eq.~\eqref{XoverScales}, indicates that this is experimentally feasible. 
This identifies this compound as a promising candidate to investigate the strong coupling between electronic and elastic degrees of freedom close to the finite-$T$ Mott endpoint and the concomitant change in universality class.

We acknowledge helpful discussions with M. Lang, I. Paul and A. Rosch. This work is supported by the DFG grants SFB 608 and FOR 960.



\begin{thebibliography}{24}
\bibitem{comment}
For the $d8$-Br crystal $\#1$ of Ref.~\cite{Bartosch10} the critical regime cannot be accessed by applying compressive pressure as the critical pressure $p_c$ is negative. However, upon substituting Br with Cl, the critical pressure can be increased, allowing for pressure tuning towards the critical regime.
\end{thebibliography}
\end{document}